\documentstyle[prd,aps,eqsecnum,preprint,tighten]{revtex}
\input epsf
\begin{document}
\preprint{\vbox to 50 pt
{\hbox{IHES/P/98/13}\hbox{CPT-98/P.3626}\vfil}}
\title{Gravitational-wave versus binary-pulsar\\
tests of strong-field gravity}
\author{Thibault Damour}
\address{Institut des Hautes \'Etudes Scientifiques, F 91440
Bures-sur-Yvette, France\\
and DARC, CNRS Observatoire de Paris, F 92195 Meudon, France}
\author{Gilles Esposito-Far\`ese}
\address{Centre de Physique Th\'eorique\cite{UPR}, CNRS Luminy, Case
907, F 13288 Marseille Cedex 9, France}
\date{Received 9 March 1998}
\maketitle

\begin{abstract}
Binary systems comprising at least one neutron star contain strong
gravitational field regions and thereby provide a testing ground for
strong-field gravity. Two types of data can be used to test the law
of gravity in compact binaries: binary pulsar observations, or
forthcoming gravitational-wave observations of inspiralling binaries.
We compare the probing power of these two types of observations
within a generic two-parameter family of tensor-scalar gravitational
theories. Our analysis generalizes previous work (by us) on
binary-pulsar tests by using a sample of realistic equations of state
for nuclear matter (instead of a polytrope), and goes beyond a
previous study (by C.M.~Will) of gravitational-wave tests by
considering more general tensor-scalar theories than the
one-parameter Jordan-Fierz-Brans-Dicke one. Finite-size effects
in tensor-scalar gravity are also discussed.
\end{abstract}
\draft

\section{Introduction}

The detection of gravitational waves by kilometric-size
laser-interferometer systems such as LIGO in the US and VIRGO in
Europe will initiate a new era in astronomy. One of the most
promising sources of gravitational waves is the inspiralling compact
binary, a binary system made of neutron stars or black holes whose
orbit decays under gravitational radiation reaction. The observation
of these systems will provide important astrophysical information,
e.g. masses of neutron stars, and direct distance measurements up to
hundreds of Mpc \cite{reviews}. It is also said that detecting
gravitational waves from inspiralling binaries should provide rich
tests of the law of relativistic gravity in situations comprising
strong-field regions (like near a neutron star or a black hole).
However, present binary-pulsar experiments already provide us with
deep and accurate tests of strong-field gravity
\cite{taylor94,twdw92,dt92}. It is therefore interesting to compare
and contrast the probing power of present (and foreseeable) pulsar
tests with that of future gravity-wave observations.

A convenient quantitative way of doing this comparison is to work
within a multi-parameter family of physically motivated (and
physically consistent) theories of gravity which differ from
Einstein's theory in their radiative and strong-field predictions.
The most natural framework of this type is the general class of
tensor-scalar theories in which gravity is mediated both by a tensor
field $g_{\mu \nu}^*$ (``Einstein metric'') and by a scalar field
$\varphi$. These theories contain one arbitrary ``coupling function''
$A(\varphi)$ defining the conformal factor relating the pure spin-2
Einstein metric $g_{\mu \nu}^*$ to the ``physical metric''
$\widetilde{g}_{\mu \nu} = A^2 (\varphi) \, g_{\mu \nu}^*$ measured
by laboratory clocks and rods. The usual Jordan-Fierz-Brans-Dicke
theory \cite{Jordan,Fierz,BransDicke} is the one parameter class of
theories defined by a coupling function $A(\varphi) = \exp
(\alpha_0 \, \varphi)$. [The coupling strength $\alpha_0$ is related
to the often used parameter $\omega$ by $\alpha_0^2 = (2\omega +
3)^{-1}$.] Will \cite{will94} has studied the quantitative
constraints on the coupling parameter $\alpha_0$ of
Jordan-Fierz-Brans-Dicke theories that could be brought by
gravitational-wave observations. His result is that in most cases the
bounds coming from gravity-wave observations will be comparable to
presently known bounds coming from solar-system
experiments (namely, $\alpha_0^2 < 10^{-3}$).
This result of Ref.~\cite{will94} seems to
suggest that gravitational-wave-based
{\it strong-field} tests of gravity do not go really beyond the
solar-system {\it weak-field} tests of gravity.
We wish, however, to emphasize that this
seemingly pessimistic conclusion is mainly due to having restricted
one's attention to the special, one-parameter Jordan-Fierz-Brans-Dicke
theory. Indeed, in this theory the strength of the coupling of the
scalar field $\varphi$ to matter is given by the constant quantity
$\alpha_0$ independently of the state of condensation of the
gravitational source. As a consequence, the predictions of the theory
differ from those of Einstein's theory by a fraction of order
$\alpha_0^2$ in all situations: weak-field ones or strong-field ones,
alike. By contrast, it has been emphasized in Refs.~\cite{def93,def96}
that the more generic tensor-scalar
theories in which the strength of the coupling of $\varphi$ to
matter, namely
\begin{equation}
\alpha (\varphi) \equiv \frac{\partial \ln A
(\varphi)}{\partial \, \varphi} \, , \label{eq1.1}
\end{equation}
depended on the value of $\varphi$, allowed for the existence of
genuine {\it strong-field effects}, by which the presence of a highly
condensed source, such as a neutron star, could generate order-unity
deviations from general relativity even if the weak-field limit of
$\alpha (\varphi)$ is arbitrarily small. [More precisely, these
non-perturbative strong-field effects take place when $\beta (\varphi)
\equiv \partial \, \alpha (\varphi) / \partial \, \varphi$ is
negative.] This led us to consider, instead of the
Jordan-Fierz-Brans-Dicke model $A(\varphi) = \exp(\alpha_0 \varphi)$,
the class of theories defined by
\begin{equation}
A(\varphi) = \exp \left( \frac{1}{2} \, \beta_0 \varphi^2 \right) \, .
\label{eq1.2}
\end{equation}

This class of theories contains two arbitrary (dimensionless)
parameters: $\beta_0$ appearing in Eq.~(\ref{eq1.2}), and $\varphi_0$,
the asymptotic value of $\varphi$ at spatial infinity. [By contrast,
in Jordan-Fierz-Brans-Dicke theory, $\varphi_0$ has no observable
effects.] Equivalently,
the two parameters in these theories can be defined as being
the strength of the linear
coupling of $\varphi$ to matter in the weak-field limit $(\varphi
\approx \varphi_0)$,
\begin{equation}
\alpha_0 = \alpha (\varphi_0) = \frac{\partial \ln
A(\varphi_0)}{\partial \, \varphi_0}
= \beta_0 \varphi_0 \, , \label{eq1.3}
\end{equation}
and the non-linear coupling parameter
\begin{equation}
\beta_0 = \frac{\partial \, \alpha (\varphi_0)}{\partial \, \varphi_0}
= \frac{\partial^2 \ln A(\varphi_0)}{\partial \, \varphi_0^2}
\, . \label{eq1.4}
\end{equation}

A convenient feature of the two-parameter family of theories
(\ref{eq1.2}) is that they have just the amount of generality needed
both to parametrize the most general boost-invariant weak-field
(``post-Newtonian'') deviations from Einstein's theory, and to
encompass nonperturbative strong-field effects. Indeed,
on the one hand, the two
theory-parameters $(\alpha_0 , \beta_0)$ determine the two well-known
Eddington-Nordtvedt-Will parameters,
\begin{equation}
\overline{\gamma} \equiv \gamma_{\rm Eddington} - 1 = -2 \, \alpha_0^2
/ (1 + \alpha_0^2) \, , \label{eq1.5}
\end{equation}
\begin{equation}
\overline{\beta} \equiv \beta_{\rm Eddington} - 1 = \frac{1}{2} \,
\beta_0 \, \alpha_0^2 / (1 + \alpha_0^2)^2 \, , \label{eq1.6}
\end{equation}
which measure the most general, boost-invariant, deviations from
general relativity at the first post-Newtonian level. (See
\cite{def92,def2pn} for the generalization of the Eddington parameters
to the second post-Newtonian level.) On the other hand, when $\beta_0
\lesssim -4$ non-perturbative strong-field effects develop in the
theories defined by Eq.~(\ref{eq1.2}).

All existing gravitational experiments can be interpreted as
constraints on the two-dimensional space of theories defined by
Eq.~(\ref{eq1.2}). In other words, we can work within the common
$(\alpha_0 , \beta_0)$ plane, and consider each gravitational
experiment (be it of weak-field or strong-field nature) as defining a
certain exclusion plot within that plane. In some recent work
\cite{def96}, we have constructed such exclusion plots, as defined by
considering both solar-system experiments and binary-pulsar
experiments. The present work will generalize these exclusion plots in
several respects: (i)~we shall plot the regions of the $(\alpha_0 ,
\beta_0)$ plane probed by future gravitational-wave observations of
neutron star--neutron star, and neutron star--black hole systems,
(ii)~we shall improve our previous study of the probing power of
binary-pulsar experiments by considering a sample of realistic nuclear
equations of states (instead of the polytropic one used by us before)
and by using updated pulsar data,
and (iii)~we shall consider the individual constraints on
$\alpha_0$ and $\beta_0$ brought by the main solar-system experiments
(instead of using published combined limits on $\overline{\gamma}$ and
$\overline{\beta}$).

This paper is organized as follows: In Sec.~II we summarize our
(numerical) approach to computing the various form factors that
describe the coupling of the scalar field $\varphi$ to a neutron star.
In Sec.~III we generalize Ref.~\cite{will94} in discussing how
gravitational wave observations can give quantitative tests of
tensor-scalar gravity. In Sec.~IV we combine and compare
gravitational-wave tests with binary-pulsar tests and solar-system
ones. Our conclusions are presented in Sec.~V, while an Appendix
discusses finite-size effects in tensor-scalar gravity.

\section{Gravitational form factors of neutron stars described by
realistic equations of state}

The orbital dynamics of a binary system depends, besides the Einstein
masses of the two objects $m_A$, $m_B$, on the effective coupling
constants $\alpha_A$, $\alpha_B$, defined as
\begin{equation}
\alpha_A \equiv \frac{\partial \ln m_A}{\partial \, \varphi_0}
\, , \label{eq2.1}
\end{equation}
as well as on their scalar-field derivatives
\begin{equation}
\beta_A \equiv \frac{\partial \, \alpha_A}{\partial \, \varphi_0} \, .
\label{eq2.2}
\end{equation}
The derivatives in Eqs.~(\ref{eq2.1}), (\ref{eq2.2}) are taken for
fixed values of the baryonic mass $\overline{m}_A$. In the limit of
negligible self-gravity $\alpha_A \rightarrow \alpha_0$ and $\beta_A
\rightarrow \beta_0$, but it was shown in Refs.~\cite{def93,def96}
that (when $\beta_0$ is negative enough) the
effective coupling constant $\alpha_A$ of a neutron star can reach
values of order unity even if the weak-field coupling constant
$\alpha_0$ is extremely small. By contrast, in Jordan-Fierz-Brans-Dicke
theory $\alpha_A$ tends to zero with $\alpha_0$. Let us also note that
for a black hole $\alpha_A = 0$ because of the no-scalar-hair theorems.

In Ref.~\cite{def92} it was shown that the parameters $m_A$, $m_B$,
$\alpha_A$, $\alpha_B$, $\beta_A$, $\beta_B$ suffice to determine both
the $(v/c)^2$-accurate conservative orbital dynamics (including
periastron precession and other relativistic deformations of Keplerian
motion) and the radiation reaction effects (monopolar, dipolar and
quadrupolar effects linked to the combined emission of scalar and
tensor waves). On the other hand, the relativistic timing of binary
pulsar systems involves, besides the above effective coupling
constants, a new parameter describing the possible field dependence of
the (Einstein-frame) inertia moment $I_A$ of the pulsar
\cite{eardley,def96}. This new parameter, entering the timing formula,
is of the form $K_A^B = k_A \, \alpha_B$ where
\begin{equation}
k_A = - \frac{\partial \ln I_A}{\partial \, \varphi_0} \, .
\label{eq2.3}
\end{equation}
For a fixed value of the baryonic mass $\overline{m}_A$, the
quantities $m_A$, $\alpha_A$, $\beta_A$ and $k_A$ depend on:
(i)~the value $\varphi_0$ of the scalar field at infinity and
the theory of gravity used (hence a dependence on $\alpha_0$ and
$\beta_0$ in our case), and (ii)~the equation of state of nuclear
matter used to describe the interior of the neutron star.

In this work, we shall consider, besides the simple polytropic equation
of state that we used before,
\begin{equation}
\widetilde{\epsilon} = \widetilde{n} \, \widetilde{m}_b + \frac{K \,
\widetilde{n}_0 \, \widetilde{m}_b}{\Gamma - 1} \ \left(
\frac{\widetilde{n}}{\widetilde{n}_0} \right)^{\Gamma} \, , \
\widetilde{P} = K \, \widetilde{n}_0 \, \widetilde{m}_b \,
\left(\frac{\widetilde{n}}{\widetilde{n}_0} \right)^{\Gamma} \, ,
\label{eq2.4}
\end{equation}
\begin{equation}
\widetilde{m}_b = 1.66 \times 10^{-24} {\rm g} \ , \ \widetilde{n}_0 =
0.1 \, {\rm fm}^{-3} \ ,\ \Gamma = 2.34 \ , \ K = 0.0195 \ ,
\label{eq2.5}
\end{equation}
a sample of three (more or less) realistic equations of state. They
describe the crust by the same equation of state (due to Baym {\it et
al.} \cite{bps71}) but differ in their descriptions of the core of
the neutron star. One description (due to Pandharipande
\cite{pandha}) gives a rather soft equation of state, a second one
(due to Wiringa {\it et al.} \cite{wiringa}) is of intermediate
stiffness, while the third one (due to Haensel {\it et al.}
\cite{haensel}) corresponds to a somewhat stiff equation of state.
This sample will allow us to see the effect of the softness of the
equation of state.

We have discussed in detail in Ref.~\cite{def96} the way to compute the
effective coupling constants $m_A$, $\alpha_A$, $\beta_A$, $k_A$ in
tensor-scalar gravity. We shall not repeat this material here. Let us
only mention a modification that we brought to the procedure discussed
in Ref.~\cite{def96} for more convenience when dealing with equations
of state given in tabular form. Starting from such tables, we first
interpolate them to define the (proper) physical energy density
$\widetilde{\epsilon}$ and the (proper) physical baryon number density
$\widetilde n$ as functions of the physical pressure $\widetilde p$.
[The tilde refers here to quantities expressed in ``physical units'',
i.e., with respect to the metric $\widetilde{g}_{\mu \nu} = A^2 \,
g_{\mu \nu}^*$.] Then, we transform the field equations by using the
pressure $\widetilde p$ as a radial variable. More precisely, this
means that we consider a set of 8 first-order differential equations
for the variation with $\widetilde p$ of the variables $M$, $\nu$,
$\varphi$, $\psi$, $\rho$, $\overline M$, $\omega$ and $\varpi$. The
symbols $M , \ldots , \varpi$ have the same meaning as in
Ref.~\cite{def96}. In particular, $\rho$ is the Schwarzschild-like
radial coordinate used in Eq.~(3.1) of \cite{def96}. Note that the
quantity $\alpha_0$ is constructed at the end of the numerical work
{}from the asymptotic value of $\varphi$ by using $\alpha_0 = \beta_0
\, \varphi_0 $, Eq.~(\ref{eq1.3}). The right-hand side of the equation
giving $d\rho / d \widetilde{p}$ is the inverse of the right-hand side
of Eq.~(3.6e) of \cite{def96}, while the right-hand sides of the 7
other equations is simply obtained [from $dQ / d \widetilde{p} \equiv
(dQ / d\rho) / (d \widetilde{p} / d\rho)$] by dividing the right-hand
sides of Eqs.~(3.6a)--(3.6h) by the right-hand side of Eq.~(3.6e). A
bonus of this modified approach is that we can numerically integrate
these differential equations on a {\it known\/} range: namely, we start
integration at the center of the star with some given central value for
$\widetilde{p}$, say $\widetilde{p}_c$, and we stop at $\widetilde{p} =
0$ (surface of the star). The value of $\rho$ at the end of integration
gives the radius $R$ of the star. We have discussed in
Refs.~\cite{def93,def96} how to obtain the quantities $\overline{m}_A$,
$m_A$, $\alpha_A$, and $I_A$ from the other results of the integration.

Actually, having chosen some nuclear equation of state, the above
numerical integration yields, for each value of $\beta_0$, and for each
initial values $\widetilde{p}_c$ (central pressure) and $\varphi_c$
(central value of the scalar field), an output which consists of
($\overline{m}_A$, $m_A$, $\alpha_A$, $I_A$, $\varphi_0$), where
$\varphi_0$ is the value of $\varphi$ at spatial infinity. [We keep
only the ``positively scalar-polarized'' configurations with $\varphi_0
\, \varphi_c > 0$. Indeed, we have shown in \cite{def96} that the
solutions with $\varphi_0 \, \varphi_c < 0$ are energetically
disfavored.] Having constructed a sufficiently dense grid (in
$(\widetilde{p}_c , \varphi_c)$ space) of such numerically integrated
models, we can then compute $\beta_A$, Eq.~(\ref{eq2.2}), and $k_A$,
Eq.~(\ref{eq2.3}), by numerically approximating the
$\varphi_0$-derivatives (with fixed $\overline{m}_A$) by finite
differences constructed from four neighboring grid points. The
expression (\ref{eq2.1}) for $\alpha_A$ is used as a test of our
numerical accuracy (we appropriately densify our grid so as to ensure
that Eq.~(\ref{eq2.1}) is satisfied to better than $5\%$). Using
finally some interpolation, we end up by generating a huge, multi-entry
table which is conveniently organized as giving, for a grid of values
of $\alpha_0$ and $\beta_0$, numerical values of $(\overline{m}_A , m_A
, \alpha_A , \beta_A , I_A , k_A)$. This is a numerical approximation
to giving $m_A$, $\alpha_A$, $\beta_A$, $I_A$ and $k_A$ as functions of
$\alpha_0$, $\beta_0$ and $\overline{m}_A$. We should also mention that
we keep only stable models in our tables, i.e., we keep only
configurations corresponding to masses increasing, when the central
pressure increases, between (formally) zero and their first maximum
(corresponding to the usual concept of the maximum mass of a neutron
star, here quantitatively modified by scalar-field effects).

\section{Testing tensor-scalar gravity from gravitational-wave
observations of inspiralling binaries}

The gravitational wave signal from inspiralling binaries will be deeply
buried in the broadband noise of interferometric detectors. To detect
it, one will have to correlate the output of the detector with a
``template wave form'', i.e., an hopefully faithful copy of the actual
time evolution of the signal. As hundreds to tens of thousands of
cycles may be usefully registered in the output of the detector, this
method of matched filtering will allow one to dig deeply into the
broadband noise, under the condition that the used template wave form
be an accurate representation of the signal over that many cycles. This
means that gravitational wave observations will be very sensitive to
the evolution of the frequency and the phase of the wave.

In tensor-scalar gravity the evolution of the phase of the
gravitational wave will be different from that in Einstein's pure
tensor gravity because this phase evolution is driven by radiation
reaction, which is modified because the system can now loose energy to
scalar waves, in addition to loosing energy to tensor waves. More
precisely, the energy lost to scalar or tensor radiation at infinity
now reads, with sufficient approximation,
\begin{equation}
\dot E = {\dot E}_{\varphi}^{\rm monopole}
+ {\dot E}_{\varphi}^{\rm dipole} +
{\dot E}_{\varphi}^{\rm quadrupole} + {\dot E}_h^{\rm quadrupole} \, ,
\label{eq3.1}
\end{equation}
where the subscript denotes the field being radiated, and where the
superscript refers to the multipolar structure of the emitted
radiation. The explicit expressions for the various energy losses have
been derived for general tensor-(multi-)scalar gravity theories in
Ref.~\cite{def92} (see also \cite{willbook}). As said above, they
depend on the (Einstein) masses $m_A$, $m_B$ and on the effective
scalar coupling parameters of the two stars, $\alpha_A$, $\alpha_B$,
$\beta_A$, $\beta_B$. Denoting $M \equiv m_A + m_B$, $X_A \equiv m_A /
M$, $X_B \equiv m_B / M = 1 - X_A$, $\nu \equiv X_A X_B$, and assuming
that the orbit is circularized (eccentricity $e=0$), we find from
Eq.~(6.52) of Ref.~\cite{def92} the following evolution law for the
orbital frequency $n = 2\pi / P_{\rm orbit}$
\begin{equation}
\frac{\dot n}{n^2} = \frac{\nu}{1 + \alpha_A \alpha_B}
\ \left[ \frac{96}{5} \,
\kappa \ \left( \frac{v}{c} \right)^5 + (\alpha_A - \alpha_B)^2
\, \left( \frac{v}{c} \right)^3 + O \left( \left( \frac{v}{c}
\right)^6 \right) \right] \, , \label{eq3.2}
\end{equation}
where
\begin{equation}
v \equiv [G_* (1 + \alpha_A \alpha_B) \, M n ]^{1/3} \label{eq3.3}
\end{equation}
(in which $G_*$ denotes the ``bare'' gravitational constant entering
the action of the Einstein metric: $\int d^4 x \, g_*^{1/2} \, R(g_*) /
16 \pi \, G_*$), and
\begin{eqnarray}
\kappa &\equiv &1 + \frac{1}{6} \, (\alpha_A X_B + \alpha_B X_A)^2 +
\frac{1}{6} \, (\alpha_A - \alpha_B)
(\alpha_A X_A + \alpha_B X_B) (X_A - X_B)
\nonumber \\
&+ &\frac{5}{48} \, \frac{\alpha_A - \alpha_B}{1 + \alpha_A \alpha_B} \
(\beta_B \alpha_A X_A - \beta_A \alpha_B X_B)
+ d_2 \, (\alpha_A - \alpha_B)^2
\, . \label{eq3.4}
\end{eqnarray}
The term $(\alpha_A - \alpha_B)^2 \, (v/c)^3$ in Eq.~(\ref{eq3.2}) is
associated to the emission of scalar dipolar waves
\cite{eardley,will77,willbook,wz89}. The first term (equal to one) in
$\kappa$ comes from the emission of tensor quadrupolar waves, the
second term is due to the emission of scalar quadrupolar waves, the
remaining terms come from $O((v/c)^5)$ contributions to the scalar
dipolar waves [responsible for the $O((v/c)^3)$ term in
Eq.~(\ref{eq3.2})]. The numerical coefficient $d_2$ has not been
computed, but this is not important in the present context as it
multiplies the same term $(\alpha_A - \alpha_B)^2$ which will be more
strongly constrained by the $O((v/c)^3)$ dipolar radiation term in
Eq.~(\ref{eq3.2}). [Let us note in passing that our parameter $\kappa$
differs by a certain factor from the one used by Will \cite{will94},
namely, $\kappa_{\rm will} = (1 + \alpha_A \alpha_B)^2 \, (1 +
\alpha_0^2)^{-3} \, \kappa_{\rm here}$.]

The expression (\ref{eq3.2}) is similar to the frequency evolution
equations used by Will \cite{will94}. Our results are, however, more
general because they apply in a generic tensor-gravity theory, while
his were restricted to the Jordan-Fierz-Brans-Dicke case. On the other
hand, starting from Eq.~(\ref{eq3.2}) we can follow his analysis in
discussing the constraints that can be derived from (future)
gravitational wave observations. Let us define a tensor-scalar ``chirp
mass'' (in time units) by
\begin{equation}
{\cal M} \equiv \frac{G_*}{c^3} \ (1 + \alpha_A \alpha_B)
\, M \, \left(
\frac{\nu \, \kappa}{1 + \alpha_A \alpha_B} \right)^{3/5}
\, , \label{eq3.5}
\end{equation}
and the corresponding adimensionalized orbital frequency
\begin{equation}
u \equiv {\cal M} \, n \, . \label{eq3.6}
\end{equation}
In terms of these variables, Eq.~(\ref{eq3.2}) reads
\begin{eqnarray}
{\cal M} \, \dot u &= & \frac{96}{5} \ u^{11/3}
\biggl\{ 1 + b \, \nu^{2/5} \,
u^{-2/3} + O (b \, \nu^{-1/5} \, u^{1/3}) \nonumber \\
&&- \left( \frac{743}{336} + \frac{11}{4} \ \nu \right)
\ \nu^{-2/5} \,
u^{2/3} \, [1 + O (\alpha^2)] \nonumber \\
&&+ 4\pi \, \nu^{-3/5} \, u \, [1 + O (\alpha^2)] + O
(u^{4/3})\biggr\} ,
\label{eq3.7}
\end{eqnarray}
where we defined
\begin{equation}
b \equiv \frac{5}{96} \ \kappa^{-3/5}
\, (1 + \alpha_A \alpha_B)^{-2/5} \,
(\alpha_A - \alpha_B)^2 \, . \label{eq3.8}
\end{equation}
The zeroth-order term $\propto u^{11/3}$ in Eq.~(\ref{eq3.7}) represent
the frequency evolution due to the emission of tensor quadrupolar
waves. The other terms represent corrections to this zeroth-order
result. The terms containing a factor $b$ or a symbolic factor
$\alpha^2$ represent the effects of the radiation of scalar waves. The
main term due to the radiation of scalar waves is the $b \, \nu^{2/5}
\, u^{-2/3}$ term. It is important to note that this term (coming from
dipolar waves) contains the large factor $u^{-2/3} \sim v^{-2} \gg 1$.
This term is the main source of the constraints on tensor-scalar
gravity coming from gravity-wave observations because it has a
different dependence on $u$ than the usual general relativistic terms.
Indeed, the usual general relativistic frequency evolution is given by
the terms without $b$ or $\alpha^2$ factors, i.e., symbolically, ${\cal
M} \, \dot u \sim u^{11/3} \, [1 + u^{2/3} + u^{3/3} + u^{4/3} +
\cdots]$.

Assuming that future gravity-wave observations from inspiralling
binaries will be well matched by filters constructed from the
standard general relativistic orbital phase evolution, we wish to
quantify what constraints they could bring on the modifications to
the frequency evolution equation (\ref{eq3.7}) by the terms
proportional to $b$ or $\alpha^2$. This problem has been tackled by
Will \cite{will94} using a matched-filter analysis. His analysis can be
applied to our case, if we follow his (plausible) assumption that the
final bounds will restrict scalar-field effects to be more or less
uniformly small compared to general relativistic ones. This means that
the only essential scalar-field term that we should consider in
Eq.~(\ref{eq3.7}) is the term $b \, \nu^{2/5} \, u^{-2/3}$. All the
scalar-field effects that simply modify usual einsteinian effects can
be neglected. Furthermore, this means also that we can approximate
$\kappa$ by one, and the chirp mass (\ref{eq3.5}) by its einsteinian
limit $G_* \, c^{-3} \, M \, \nu^{3/5}$. Finally, we can translate
Will's final bound on the Jordan-Fierz-Brans-Dicke coupling parameter
$\alpha_0^2$ (his Eq.~(\ref{eq1.5})), in the following bound on $b$ or
$(\alpha_A - \alpha_B)^2$:
\begin{equation}
\frac{48}{5} \ b \approx \frac{1}{2} \, (\alpha_A - \alpha_B)^2 < 1.46
\times 10^{-5} \, \left( \frac{\cal M}{T_{\odot}} \right)^{7/3} \,
\nu^{-2/5} \, \left( \frac{10}{S/N} \right) \, , \label{eq3.9}
\end{equation}
where $T_{\odot} = G_* \, m_{\odot} \, c^{-3}$ is the gravitational
time associated to a solar mass. This bound can also be rewritten in
terms of $M = m_A + m_B$ and $\nu = m_A \, m_B / (m_A + m_B)^2$:
\begin{equation}
(\alpha_A - \alpha_B)^2 < 2.92 \times 10^{-5} \,
\left( \frac{M}{m_{\odot}}
\right)^{7/3} \, \nu \, \left( \frac{10}{S/N} \right)
\, . \label{eq3.10}
\end{equation}
In Eqs.~(\ref{eq3.9}) and (\ref{eq3.10}), $S/N$ denotes the
signal-to-noise ratio (after applying a matched filter) of a
LIGO/VIRGO-type detector. [The analysis of Ref.~\cite{will94} assumes
the noise spectral density of a LIGO ``advanced detector''.]

We have focussed above on the indirect effect of scalar-wave emission
on the phase evolution of tensor waves for the following reason.
Relatively to the main tensor-quadrupolar contribution, this effect
is of order $(\alpha_A - \alpha_B)^2 \, (c/v)^2 \gg (\alpha_A -
\alpha_B)^2$. Moreover, because of possible nonperturbative
scalar-field effects associated to the strong self-gravity of neutron
stars $(\alpha_A - \alpha_B)^2$ can reach values of order unity, even
if the weak-field scalar coupling $\alpha_0 \ll 1$. By contrast, the
more direct effect of scalar-wave emission on the differential forces
acting on any (local) detector of gravitational waves can be described
by the transverse projection of the physical metric $\widetilde{g}_{\mu
\nu}$
\begin{equation}
\widetilde{g}_{ij}^T = (A^2 (\varphi_0 + \phi) \, g_{ij}^*)^T \approx
h_{ij}^{*TT} + 2 \, \alpha_0 \, \phi \, \delta_{ij}^T
\, , \label{eq3.11}
\end{equation}
where $\delta_{ij}^T \equiv \delta_{ij} - n_i \, n_j$ is the projection
of $\delta_{ij}$ transversally to the direction $n^i$ of propagation of
the wave (see, e.g., \cite{def92}), and where we choose units such that
$A (\varphi_0) = 1$. The main direct effect of scalar-wave emission,
associated, to scalar dipolar emission, is then (see Eq.~(6.33) of
\cite{def92})
\begin{equation}
2 \, \alpha_0 \, \phi^{\rm dipole} \, \delta_{ij}^T =
- \frac{2}{r} \ \alpha_0
\, (\alpha_A - \alpha_B) \ \frac{m_A m_B}{M} \ n_i (z_A^i - z_B^i) \, .
\label{eq3.12}
\end{equation}
Compared to the standard tensor-quadrupolar wave, this is of order
$\alpha_0 \, (\alpha_A - \alpha_B) \ (c/v)^2$. Even when
nonperturbative strong-field effects generate $\alpha_A - \alpha_B =
O(1)$, we see that the direct effects of scalar-wave emission are down
by a factor $\alpha_0 \ll 1$ compared to the indirect ones. One
therefore {\it a priori\/} expects that these direct effects will be
less sensitive probes
of strong-field deviations from general relativity than
the indirect ones considered above. This qualitative-argument-based
conclusion is confirmed by the results of Will (Sec.~IV of
\cite{will94}) who quantified the lower sensitivity of matched
filtering to amplitudes, as compared to phases.

Let us also note that Refs. \cite{def93,def96} have emphasized
that gravitational collapse and neutron-star binary coalescence (by
contrast with inspiral) might exhibit observably significant direct and
indirect scalar-wave effects due to the combined possibilities of
strong-scalar-field effects and {\it monopolar\/} radiation. [However,
the recent investigation of Ref.~\cite{novak} suggests that these
effects will be too small to be of observational interest.]

\section{Combining gravity-wave, pulsar and solar-system tests}

Thanks to our consideration of a two-dimensional ``mini-space'' of
alternative gravity theories, we can combine various experimental tests
of relativistic gravity, and compare their probing power. Indeed, each
experimental constraint separates the $(\alpha_0 , \beta_0)$ plane into
two regions: an allowed one and an excluded one (at some fixed
confidence level). We can therefore compare the probing power of
various experimental tests by drawing a combined exclusion plot in the
$(\alpha_0 , \beta_0)$ plane.

\begin{figure}
\begin{center}\leavevmode\epsfbox{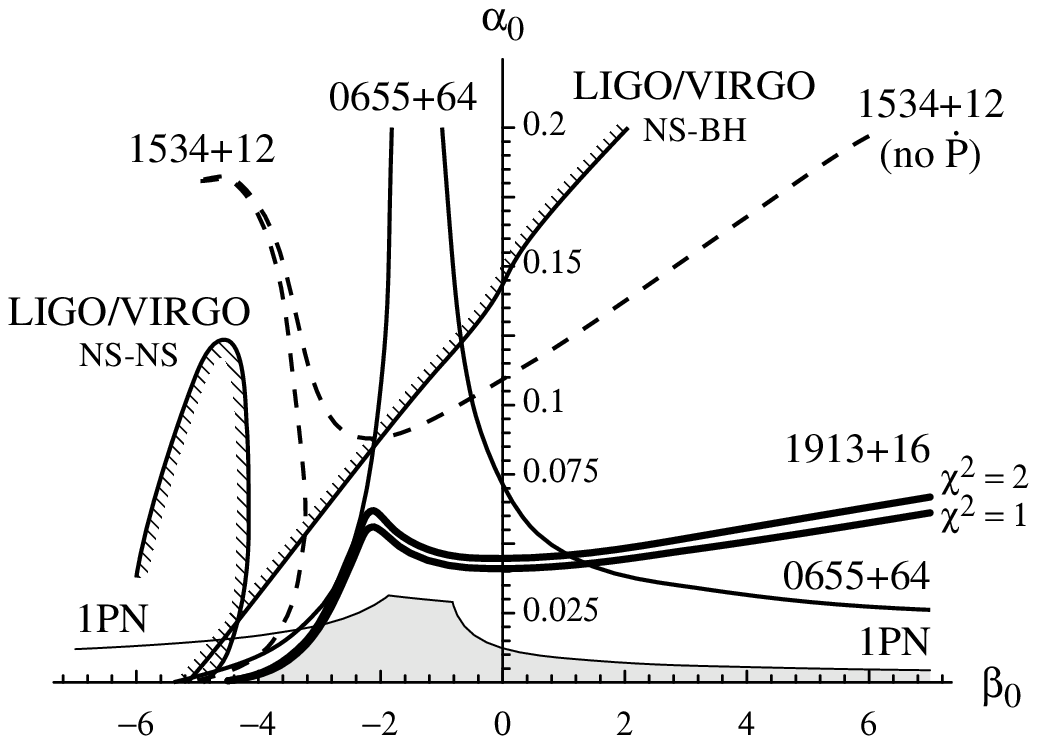}\end{center}\vskip 1pc
\caption{Region of the $(\alpha_0,\beta_0)$ theory plane allowed by
solar-system tests, binary-pulsar experiments, and future
gravity-wave detections, in the case where nuclear matter is
described by the polytrope
(\protect{\ref{eq2.4}})-(\protect{\ref{eq2.5}}). In view of the
reflection symmetry $\alpha_0\rightarrow -\alpha_0$, we plot only the
upper half plane. The region allowed by solar-system tests is below
the thin line labeled ``1PN''. The PSR 0655+64 data constrain the
values of $\alpha_0$ and $\beta_0$ to be between the two solid lines.
The regions allowed by the PSR 1913+16 and PSR 1534+12 tests lie
respectively to the right of the bold line and of the the dashed
line. The horn-shaped region at the top-left of the dashed line is
removed if the observable $\dot P_b^{\rm obs}$ is taken into account
for PSR 1534+12. Each of these curves determines the level $\chi^2 =
1$ for the corresponding test. We have also plotted the level
$\chi^2 = 2$ for PSR 1913+16 to underline that the precise value of
$\chi^2$ is not very significant in the region where binary-pulsar
experiments are more constraining than solar-system tests
($\beta_0\,\protect{\lesssim}-3$). The regions excluded by the
gravity-wave observation limit (\protect{\ref{eq3.10}}), with a
signal-to-noise ratio $S/N = 10$, lie on the hatched
sides of the curves labeled ``LIGO/VIRGO''. The case of a
$1.4\,m_\odot$-neutron-star and a $10\,m_\odot$-black-hole binary
system is labeled ``NS-BH'', whereas the case of a
1913+16-like binary-neutron-star system is labeled ``NS-NS''.
The region simultaneously allowed by all the tests is shaded.}

\label{fig1}
\end{figure}

\begin{figure}
\begin{center}\leavevmode\epsfbox{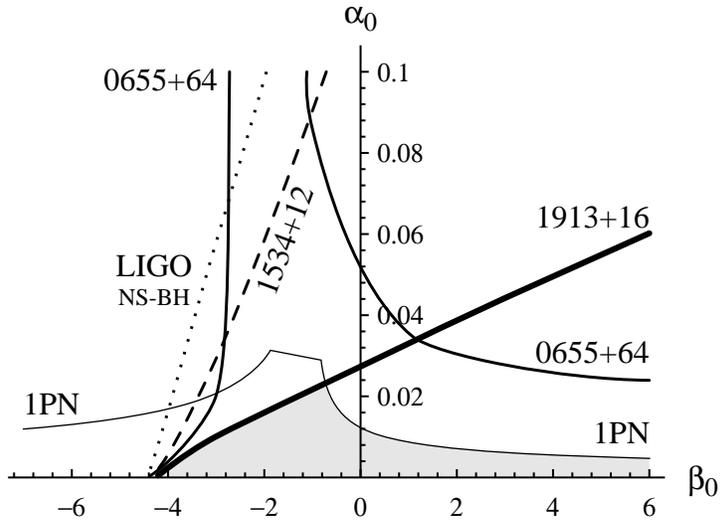}\end{center}\vskip 1pc
\caption{Same plot as Fig.~1 in the case of a soft equation of state
(Pandharipande). The region possibly excluded by the LIGO/VIRGO
detection of a ($1.4\,m_\odot$) neutron star--($10\,m_\odot$) black
hole system lies above the dotted line. The bubble-like region at the
left of Fig.~1 (binary-neutron-star system detected by LIGO/VIRGO)
does not exist in the case of this soft equation of state.}
\label{fig2}
\end{figure}

\begin{figure}
\begin{center}\leavevmode\epsfbox{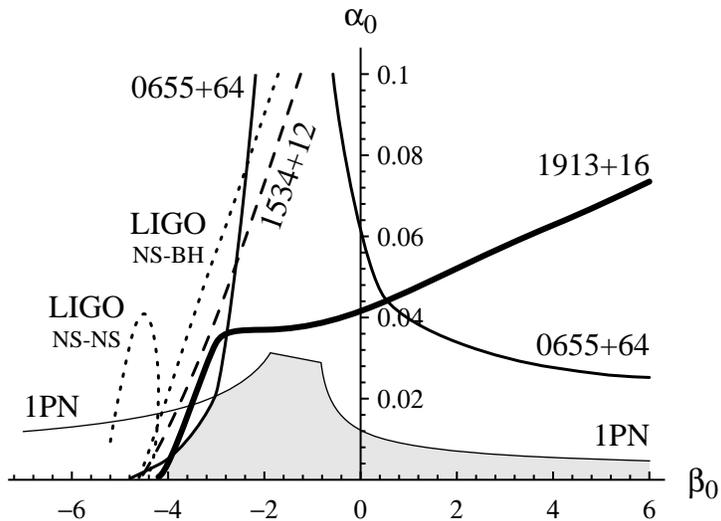}\end{center}\vskip 1pc
\caption{Same plot as Fig.~2 for a medium equation of state
(Wiringa {\it et al.}). Dotted lines indicate the regions excluded by
future gravitational-wave observations, respectively inside the bubble
for the NS-NS case, and above the straight line for the NS-BH case.}
\label{fig3}
\end{figure}

\begin{figure}
\begin{center}\leavevmode\epsfbox{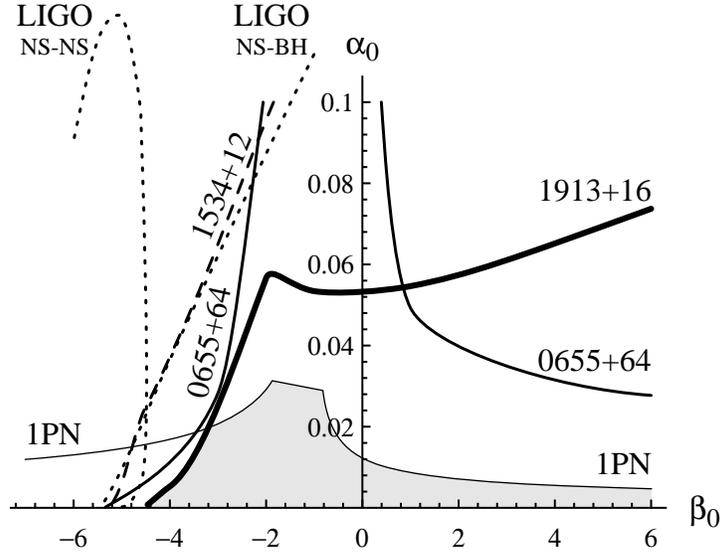}\end{center}\vskip 1pc
\caption{Same plot as Fig.~3 for a stiff equation of state (Haensel
{\it et al.}). Note that the bubble excluded by the detection of a
binary-neutron-star system by LIGO/VIRGO is much larger when the
equation of state is stiff.}
\label{fig4}
\end{figure}

We shall consider three solar-system tests of weak-field gravity. The
measurement of the Shapiro time delay by the Viking mission
\cite{reasenbergetal} as well a some VLBI measurements
\cite{robertson,lebach} yield the $1
\sigma$ bound $\vert \overline{\gamma} \vert < 2 \times 10^{-3}$, which
(in view of Eq.~(\ref{eq1.5})) gives
\begin{equation}
\alpha_0^2 < 10^{-3} \quad (1 \sigma) \, . \label{eq4.1}
\end{equation}
The measurement of the perihelion advance of Mercury \cite{shapiro90}
gives $\frac{1}{3}\ \vert 2 \, \overline{\gamma} - \overline{\beta}
\vert < 10^{-3}$ which, using Eqs.~(\ref{eq1.5}) and (\ref{eq1.6}),
translates into
\begin{equation}
\left\vert \alpha_0^2 + \frac{1}{8} \ \beta_0 \,
\alpha_0^2 \right\vert < 7.5\times 10^{-4}
\quad (1 \sigma) \, . \label{eq4.2}
\end{equation}
Finally, the Lunar Laser Ranging experiment \cite{LLR} yields
$-1.7\times 10^{-3}< 4 \, \overline{\beta} - \overline{\gamma} <
3\times 10^{-4}$ which translates into
\begin{equation}
-8.5\times 10^{-4} < \alpha_0^2 + \beta_0 \, \alpha_0^2 < 1.5\times
10^{-4} \quad (1 \sigma) \, .
\label{eq4.3}
\end{equation}
These three constraints, taken together, exclude the region above the
solid line labeled ``1PN'' in Figs.~\ref{fig1}--\ref{fig4}. Concerning
binary-pulsar experiments we shall take into account three of them.

The PSR 1913+16 experiment uses the values of three
well-measured phenomenological timing observables (``post-Keplerian
parameters'' \cite{dt92}) $\dot{\omega}^{\rm obs}$, $\gamma^{\rm obs}$
and $\dot{P}_b^{\rm obs}$. [We neglect the low-precision measurement
of the post-Keplerian parameters $r$ and $s$.] Here
$\dot{\omega}^{\rm obs}$ denotes the observed secular rate of advance
of the periastron, $\gamma^{\rm obs}$ denotes the observed value of
the ``Einstein'' time-dilation parameter (entering $\Delta_E =
\gamma \sin u$ \cite{dt92}), and $\dot{P}_b^{\rm obs}$ denotes
the secular change of the orbital period. The values we shall take
for these observed parameters are \cite{taylorCQG}
\begin{mathletters}
\label{eq4.4}
\begin{eqnarray}
\dot{\omega}^{\rm obs} &= &4.226 \, 621 \, (11)^\circ\ {\rm yr}^{-1}
\, ,
\label{eq4.4a} \\
\gamma^{\rm obs} &= &4.295 \, (2) \times 10^{-3} \ {\rm s}
\, , \label{eq4.4b}
\\
\dot{P}_b^{\rm obs} - \dot{P}_b^{\rm gal} &=
&-2.4101(85)\times 10^{-12} \, , \label{eq4.4c}
\end{eqnarray}
\end{mathletters}
where figures in parentheses represent (realistic) $1\sigma$
uncertainties in the last quoted digits. In Eq.~(\ref{eq4.4c}),
$\dot{P}_b^{\rm gal}$ represents the sum of various galactic effects
which must be subtracted from $\dot{P}_b^{\rm obs}$ to be able to
compare it to theoretical predictions \cite{dt91}. [We will take
into account below the small modifications brought to it by
tensor-scalar gravity, Eq.~(9.22) of \cite{def92}.]
We need also the values of the Keplerian parameters
\begin{mathletters}
\label{eq4.5}
\begin{eqnarray}
P_b &= &27 \, 906.980 \, 780 \, 4 (6) \ {\rm s} \, , \label{eq4.5a} \\
e &= &0.617 \, 130 \, 8 (4) \, . \label{eq4.5b}
\end{eqnarray}
\end{mathletters}
As explained in detail in Ref.~\cite{def96} one converts the three
phenomenological (theory-independent) measurements (\ref{eq4.4}) into
constraints on alternative gravity theories by using the predictions
that, say, tensor-scalar gravity make for $\dot{\omega}$, $\gamma$,
$\dot{P}_b$ as functions of the two ({\it a priori\/} unknown) masses
$m_A$, $m_B$ and of the parameters (here $\alpha_0$, $\beta_0$)
defining each tensor-scalar theory. As already mentioned above, the
dependence on the theory-parameters $\alpha_0$, $\beta_0$ goes through
the dependence of the post-Keplerian parameters $\dot{\omega}$,
$\gamma$ and $\dot{P}_b$ on the effective coupling constants
$\alpha_A$, $\alpha_B$, $\beta_A$, $\beta_B$, $k_A$. The explicit
expressions of $\dot{\omega}^{\rm theory} (m_A , m_B , \alpha_A ,
\beta_A , k_A , \ldots)$, $\gamma^{\rm theory} (m_A , m_B , \alpha_A ,
\beta_A , k_A , \ldots)$ and $\dot{P}_b^{\rm theory} (m_A , m_B ,
\alpha_A , \beta_A , k_A , \ldots)$ have been written down in
Refs.~\cite{def92} and \cite{def96}. We shall not reproduce them here.
Using these expressions, we then define a goodness-of-fit statistics by
minimizing over the physically {\it a priori\/} undetermined masses:
\begin{equation}
\chi^2 \, (p_i^{\rm obs} ; \alpha_0 , \beta_0) =
{\rm min}_{m_A , m_B} \,
\chi^2 \, (p_i^{\rm obs} ; m_A , m_B , \alpha_A ,
\beta_A , k_A , \ldots ) \, ,
\label{eq4.6}
\end{equation}
where, for the simultaneous measurement of any number of post-Keplerian
parameters $p_i$, one would define
\begin{equation}
\chi^2 \, (p_i^{\rm obs} ; m_A , m_B , \alpha_A ,
\beta_A , \ldots ) \equiv
\sum_i \left(\sigma_{p_i}^{\rm obs}\right)^{-2} \left(p_i^{\rm
theory} (m_A , m_B ,
\alpha_A , \beta_A , \ldots ) - p_i^{\rm obs}\right)^2 \, .
\label{eq4.7}
\end{equation}
Here, we neglect the correlations between the measurements of the
various post-Keplerian parameters.

In principle, when considering one or more pulsar experiments
simultaneously, one can associate a confidence level (in the bayesian
sense) to any region ${\cal R}$ in the theory-plane by integrating over
${\cal R}$ the normalized {\it a posteriori\/} probability density
\begin{equation}
d\alpha_0 \, d\beta_0 \, \pi_{\rm post} (\alpha_0 , \beta_0) =
d\alpha_0 \,
d\beta_0 \, N \, \pi_{\rm prior} (\alpha_0 , \beta_0) \,
W_1 ({\bf p}_1^{\rm
obs} \mid \alpha_0 , \beta_0) \, W_2 ({\bf p}_2^{\rm obs}
\mid \alpha_0 ,
\beta_0) \ldots \label{eq4.8}
\end{equation}
where the probability (density) $W_1$ to observe, in the first pulsar
experiment, the multiplet ${\bf p}_1^{\rm obs}$ (e.g.
$(\dot{\omega}^{\rm obs} , \gamma^{\rm obs} , \dot{P}_b^{\rm obs})$)
of post-Keplerian parameters can be approximated by the Gaussian
\begin{equation}
W_1 ({\bf p}_1^{\rm obs} \mid \alpha_0 , \beta_0) = \exp \left[ -
\frac{1}{2} \ \chi^2
({\bf p}_1^{\rm obs} ; \alpha_0 , \beta_0)\right] \label{eq4.9}
\end{equation}
with the definition (\ref{eq4.6}) above. In Eq.~(\ref{eq4.8}), $N$ is a
normalization constant (such that $\int \int d\alpha_0 \, d\beta_0 \,
\pi_{\rm post} (\alpha_0 , \beta_0) = 1$) and $\pi_{\rm prior}
(\alpha_0 , \beta_0)$ is some {\it a priori\/} probability density for
the values of the theory parameters. A common (and convenient) way of
choosing some confidence regions ${\cal R}$ is to define them from the
various $\chi^2$. One can consider either the overlap (assumed to
exist) of the various individual $\chi^2$ contour levels $\chi^2 ({\bf
p}_1^{\rm obs} ; \alpha_0 , \beta_0) = k_1$, $\chi^2 ({\bf p}_2^{\rm
obs} ; \alpha_0 , \beta_0) = k_2$, etc., or the combined $\chi^2$
contour levels: $\chi_{\rm tot}^2 (\alpha_0 , \beta_0) = k_{\rm tot}$,
with $\chi_{\rm tot}^2 \equiv \sum_a \chi^2 ({\bf p}_a^{\rm obs} ;
\alpha_0 , \beta_0)$. To each choice of ${\cal R}$, we associate a
bayesian probability (``probability of causes''; the causes being here
the unknown theory parameters $\alpha_0 , \beta_0$) by integrating
(\ref{eq4.8}) over ${\cal R}$. Here, we shall not try to quantify
confidence levels in this way. We shall content ourselves by plotting
individual $\chi^2 = 1$ contours. If several independent so-defined
pulsar contours overlap, it is clear that the overlap region will
define a rather high confidence level region for the total experiment
combining these independent pulsar measurements.

As discussed in Section II above, it is essential in this discussion
to decide upon some equation of state for nuclear matter. Each choice
of nuclear equation of state defines its own corresponding exclusion
plot. For example, we expect soft equations of state (like
Pandharipande's) to yield stronger constraints on the theory
parameters $(\alpha_0 , \beta_0)$. Indeed, they lead to more highly
condensed neutron star models, and thereby to generically higher
values of the effective couplings $\alpha_A$, $\beta_A$, $k_A$ for
given $(\alpha_0 , \beta_0)$ \cite{def92}. We shall consider
successively the following equations of state: (i) in Fig.~\ref{fig1}
the polytrope (\ref{eq2.4})-(\ref{eq2.5}) (as used in our previous
work \cite{def96}), (ii) in Fig.~\ref{fig2} the (soft) Pandharipande
equation of state, (iii) in Fig.~\ref{fig3} the (medium) Wiringa {\it
et al.} equation of state, and (iv) in Fig.~\ref{fig4} the (stiff)
Haensel {\it et al.} equation of state.

The PSR 1534+12 experiment consists mainly of the high-precision
measurements of three post-Keplerian parameters (which have to deal
with strong-field effects without mixing of radiative ones
\cite{twdw92}): $\dot{\omega}^{\rm obs}$, $\gamma^{\rm obs}$ and
$s^{\rm obs}$. Here $s^{\rm obs}$ denotes a phenomenological parameter
measuring the shape of the gravitational time delay
\cite{dd86,dt92}. The values we shall take for these three observable
parameters are \cite{saclnttw97}
\begin{mathletters}
\label{eq4.10}
\begin{eqnarray}
\dot{\omega}^{\rm obs} &= & 1.755\,76(4)\,{}^\circ\ {\rm yr}^{-1}\, ,
\label{eq4.10a} \\
\gamma^{\rm obs} &= & 2.066(10)\times 10^{-3}\ {\rm s} \, ,
\label{eq4.10b} \\
s^{\rm obs} &= & 0.982(7)\, . \label{eq4.10c}
\end{eqnarray}
\end{mathletters}
We shall also take into account the lower-precision measurement of the
range-of-time-delay parameter $r^{\rm obs}$
\begin{equation}
r^{\rm obs} = 6.7(1.3) \times 10^{-6}\ {\rm s}\, ,
\label{eq4.11}
\end{equation}
and we shall need the Keplerian observables
\begin{mathletters}
\label{eq4.12}
\begin{eqnarray}
P_b &= &36\,351.702\,6587(35) \ {\rm s} \, , \label{eq4.12a} \\
e &= &0.273\,677\,7(5) \, , \label{eq4.12b} \\
x &= &3.729\,463(3) \ {\rm s} \, . \label{eq4.12c}
\end{eqnarray}
\end{mathletters}
Reference \cite{saclnttw97} gives also the observed value of the
radiative parameter $\dot{P}_b^{\rm obs}$. However, it underlines that
the theoretical significance of this measurement is (at present) highly
uncertain because of the lack of an independent, reliable measurement
of the distance to PSR 1534+12. Indeed, as discussed in \cite{dt91},
several galactic effects have to be subtracted from $\dot{P}_b^{\rm
obs}$ before equating it to the theoretical prediction $\dot{P}_b^{\rm
theory}$. These corrections are relatively small and sufficiently well
known in the case of PSR 1913+16, while they are relatively large and
insufficiently known in the case of PSR 1534+12. Because of this
insufficient knowledge of the theoretically relevant combination
$\dot{P}_b^{\rm obs} - \dot{P}_b^{\rm gal}$, we shall not take into
account this observable in our exclusion plots below. [An alternative
choice would be to take it into account but to enlarge the errors
induced by the uncertainty on the distance. In that case, we found that
its main effect is to forbid the horn-shaped region at the top-left of
Fig.~\ref{fig1}, which is anyway already ruled out by solar-system
experiments as well as other binary-pulsar tests.]

Finally, we follow Ref.~\cite{def96} in taking also into account the
data on PSR 0655+64. This binary pulsar is composed of a neutron star
of mass $m_A \approx 1.4 \, m_{\odot}$ and a white dwarf companion of
mass $m_B \approx 0.8 \, m_{\odot}$, moving around each other on a
nearly circular orbit in a period of about one day. This dissymmetric
system is, potentially, a strong emitter of scalar dipolar waves.
Indeed, we saw above that dipolar radiation losses are proportional
to $(\alpha_A - \alpha_B)^2 \, (v/c)^3$. Here $\alpha_B$ does not
differ significantly from the weak-field coupling $\alpha_0$ because
the self-gravity of a white dwarf is very small, while $\alpha_A$ can
reach values of order unity. We refer to Ref.~\cite{def96} for the
discussion of the constraints obtainable from this system. The end
result is that we get the following (conservative) $1\sigma$
constraint
\begin{equation}
[ \alpha_A (m_A) - \alpha_0]^2 < 3 \times 10^{-4} \, . \label{eq4.13}
\end{equation}
We use the corresponding $\chi_{0655+64}^2 = 1$ level in our exclusion
plots.

For all the equations of state that we consider, we see on the
exclusion plots Fig.~\ref{fig1}--\ref{fig4} that the most stringent
constraints coming from pulsar experiments are obtained by combining
the 1913+16 exclusion region for $\beta_0 \lesssim + 1$ with that from
0655+64 for $\beta_0 \gtrsim + 1$. The resulting theoretical bounds
are somewhat less stringent (by a factor of a few) than solar-system
bounds when $\beta_0 \gtrsim \beta_c$, while for $\beta_0 \lesssim
\beta_c$ pulsar experiments essentially exclude an infinite domain of
the $(\alpha_0 , \beta_0)$ plane which remained allowed by
solar-system experiments. Here, $\beta_c$ denotes the (negative)
critical value of $\beta_0$ below which nonperturbative strong-field
effects develop, thereby exhibiting the unique strong-field probing
power of pulsar experiments. As we see on
Figs.~\ref{fig1}--\ref{fig4}, the value of $\beta_c$ depends on the
equation of state. In particular, a soft equation of state leads to
highly condensed neutron star configurations and, thereby, develop
nonperturbative effects earlier than stiff equations of state: in
other words $-\beta_c^{\rm soft} < - \beta_c^{\rm stiff}$. This is
visible on Fig.~\ref{fig2} (Pandharipande) where the pulsar bound is
more stringent than the solar-system ones for $\beta_0 \lesssim -1$,
while for stiffer equation of state it becomes more stringent only
when $\beta_0 \lesssim -3$.

Finally, we have added, for comparison, the exclusion regions defined
by the gravity-wave observation limit (\ref{eq3.10}), assuming a
signal-to-noise ratio $S/N = 10$. In absence of gravity-wave
observations telling us about the precise masses of real inspiralling
binaries, we have considered two fiducial cases: (i)~a
two-neutron-star system with Einstein masses $m_A =1.441 \,
m_{\odot}$, $m_B = 1.388\, m_{\odot}$ (as measured in PSR
1913+16 when interpreted in general relativity), and (ii)~a neutron
star--black hole system with $m_A = 1.4\, m_{\odot}$, $m_B = 10 \,
m_{\odot}$. In case (i) neither the precise numerical values of the
masses nor the fact that we fix Einstein masses instead of baryonic
masses is crucial. What is crucial in our definition of the fiducial
case (i) is that we assume a {\it fractional\/} mass difference
$\Delta m / m \approx \Delta \overline m / \overline m \approx 4\%$,
as large as in PSR 1913+16. Indeed, the over important dipolar
radiation is proportional to $[\alpha_A (m_A) - \alpha_B (m_B)]^2
\propto (\Delta m / m)^2$ as $\Delta m / m \rightarrow 0$. In case
(ii), the no-scalar-hair theorems guarantee that $\alpha_B = 0$ for a
black hole (see, e.g., \cite{def92}), so that neutron-star-black-hole
systems are always good {\it a priori\/} probes of possible scalar
dipolar radiation.

Because of the complexity of the numerical calculation of the
strong-field form factors of neutron stars (see Sec.~II above), we
could compute more precisely the exclusion regions for the polytropic
equation of state, Eqs.~(\ref{eq2.4})-(\ref{eq2.5}). Therefore, it is
in Fig.~\ref{fig1} that one sees best the shape of the regions
possibly excluded\footnote{We assume here that general relativity is
the (nearly) exact description of gravity chosen by nature, and we
discuss constraints on deviations away from general relativity.} by
future gravitational wave data. The fiducial case (i) (\`a la 1913+16)
excludes an ellipsoidal bubble which touches the $\beta_0$ axis around
$\beta_0 \approx -5$, while the fiducial case (ii) (neutron
star--black hole) excludes the region {\it above\/} the nearly
straight line $\alpha_0 - 0.03 \, \beta_0 \approx 0.15$ represented in
Fig.~\ref{fig1}. The corresponding excluded regions for realistic
equations of state can be recognized on Figs.~\ref{fig2}--\ref{fig4}
as deformations of the just described regions for the polytropic
case. Note that the bubble excluded by 1913+16-like systems (i)
is smaller when the EOS is softer, and that it even disappears for
Pandharipande's equation of state (Fig.~\ref{fig2}). In that case,
the detection of such a system by a gravitational-wave interferometer
would not constrain at all the space of theories.

The value $S/N = 10$ chosen for Figs.~\ref{fig1}--\ref{fig4}
corresponds to the conventional event rate of 3 binary-neutron-star
coalescences per year in a radius of 200 Mpc, and to a probably
smaller event rate for neutron-star-black-hole coalescences. However,
as pointed out by Will \cite{willcomm}, the event rate is
only slightly relevant to our discussion since a single system can
suffice to constrain the space of gravity theories. It is therefore
interesting to consider also the lucky discovery of an exceptionally
near system, with a signal-to-noise ratio as large as $S/N = 100$.
The corresponding exclusion plots are displayed in Fig.~\ref{fig5},
for the same polytropic equation of state as in Fig.~\ref{fig1}. The
bubble excluded by the 1913+16-like system is much larger, but still
not competitive with present binary-pulsar tests. Similarly, the
neutron-star-black-hole system is more constraining than in
Fig.~\ref{fig1}, but the slope of the corresponding line is only
reduced by a factor $\sim \sqrt{10}$, although the signal-to-noise
ratio is 10 times larger. This is due to the fact that the dominant
dipolar radiation is proportional to the square of $\alpha_A$.
Therefore, the lucky discovery of a nearby neutron-star-black-hole
system would be slightly more constraining than present binary-pulsar
tests in the region $-3 \lesssim \beta_0 < 0$, but not better than
present solar-system experiments. Moreover, one must keep in mind
that solar-system experiments will improve in the mean time. In
particular, NASA's Gravity Probe B mission (due for launch in 2000)
is expected to improve the probing of $\alpha_0$ down to the level
$\alpha_0 \sim \sqrt{10^{-5}} \approx 3\times 10^{-3}$.

Since we are mentioning the possibility of lucky discoveries, like
those of PSRs 1913+16 and 1534+12, let us also quote the constraints
which could be achieved if a binary pulsar with a black-hole
companion were discovered. In that case, the main dipolar radiation
would be proportional to $\alpha_A^2$ (since $\alpha_B = 0$ for a
black hole), instead of the small factor $(\alpha_A-\alpha_B)^2
\propto (\Delta m /m)^2$ appearing for binary-neutron-star systems.
The mass of the black hole is not a crucial parameter for this
discussion (we take $m_B = 10\, m_\odot$). Assuming an orbital
period $P_b$ and a measurement accuracy for $\dot P_b$ similar to
those of PSR 1913+16, one finds that the constraints on $\alpha_0$
would be tightened by a factor\footnote{This improvement factor of
80 on $\alpha_0$ comes mainly from the ratio $|\alpha_{\rm NS} -
\alpha_{\rm BH}|/|\alpha_A - \alpha_B| \approx |c_{\rm NS} -
c_{\rm BH}|/|c_A - c_B|$, where the ``compactness'' parameters
\cite{def92} for neutron stars and black holes are respectively $c_A
\approx 0.21\, m_A/m_\odot$ and $c_{\rm BH} = 1$. (Here $m_A =
1.441\,m_\odot$, $m_B = 1.388\, m_\odot$, $m_{\rm NS} \approx
1.4\,m_\odot$, and the value of $m_{\rm BH}$ does not matter.)} $\sim
80$, {\it i.e.}, that the bold lines of Figs.~\ref{fig1}--\ref{fig5}
would cross the vertical axis around $\alpha_0 \approx 0.05/80
\approx 6\times 10^{-4}$. In terms of the Eddington parameter
$\gamma_{\rm Edd}$, Eq.~(\ref{eq1.5}), this corresponds to a level
$|\gamma_{\rm Edd} - 1| \lesssim 10^{-6}$, which is about a thousand
times tighter than present solar-system limits, and ten times better
than the probing level expected from Gravity Probe B. This underlines
that, in the future, binary-pulsar tests may become competitive with,
or even supersede, solar-system experiments even in the region
$\beta_0 > -4$ of the theory plane.

\section{Conclusions}

\begin{figure}
\begin{center}\leavevmode\epsfbox{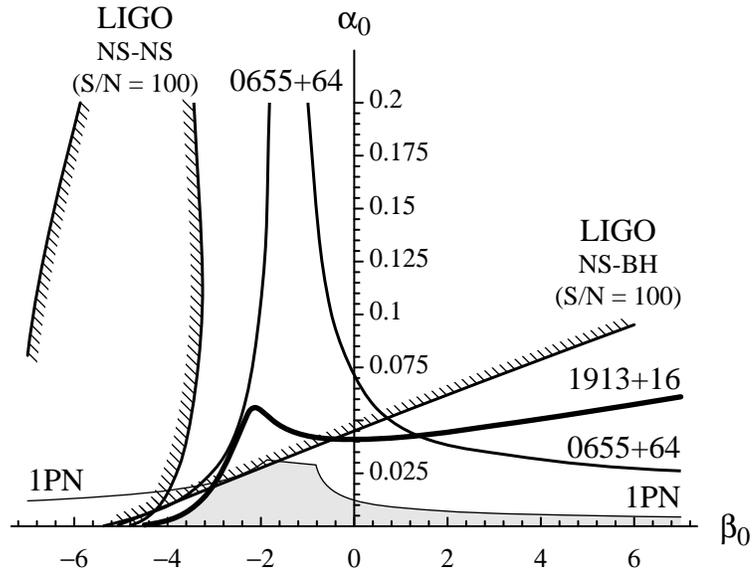}\end{center}\vskip 1pc
\caption{Same plot as Fig.~1, assuming the same polytropic
equation of state, but a signal-to-noise ratio $S/N = 100$ for
the LIGO (hatched) curves. For clarity, the dashed line
corresponding to the PSR 1534+12 test has been suppressed.}
\label{fig5}
\end{figure}

The main conclusion of the comparison carried out in
Figs.~\ref{fig1}--\ref{fig5} is that, in all cases, future LIGO/VIRGO
observations of inspiralling compact binaries turn out not to be
competitive with present binary-pulsar tests in their {\it
discriminating\/} probing of the strong-field, and radiative, aspects
of relativistic gravity. This conclusion may seem paradoxical. It
should not be interpreted negatively against LIGO/VIRGO observations
which, as shown in Figs.~\ref{fig1}--\ref{fig5}, will independently
probe strong-field gravity and will exclude regions of parameter
space allowed by solar-system experiments. Rather, it is simply a
reminder that binary-pulsar experiments are superb tools for probing
strong-field and radiative aspects of gravity. It is also somewhat a
good news for gravitational wave data analysis (which promises to be
already a very challenging task even if one {\it a priori\/} assumes
the validity of general relativity; see, e.g., \cite{dis98}). Indeed,
our results Figs.~\ref{fig1}--\ref{fig5} indicate that our present
experimental knowledge of the law of relativistic gravity is
sufficient to justify using general relativity as the standard theory
of gravitational radiation.

Note that this conclusion explicitly refers to the quantitative
probing of plausible\footnote{Within the presently accepted framework
for low-energy fundamental physics, namely field theory, the only
alternative (long-range) gravity theories which, (i) do not violate
the basic tenets of field theory, and (ii) are not already
necessarily extremely constrained by existing equivalence-principle
tests, are the tensor-scalar gravity theories.} {\it deviations\/}
{}from general relativity. At the {\it qualitative\/} level, and also
at the {\it non-discriminating\/} quantitative level, LIGO/VIRGO
observations will bring invaluable advances in our experimental
knowledge of relativistic gravity. First, they will provide the first
direct observation of gravitational waves far in the wave zone (while
binary pulsar experiments prove the reality of the propagation with
finite velocity of the gravitational interaction in the near zone of
a binary system). Second, they will (hopefully) lead to superb
additional confirmations of general relativity through the
observation of the wave forms emitted during the inspiral and
coalescence of neutron stars or black holes; see, e.g.,
\cite{reviews,bs9495}.

Independently of the comparison between LIGO/VIRGO tests and
binary-pulsar tests, the present paper has provided the first
systematic study of the influence of the nuclear equation of state on
the theoretical probing power of binary-pulsar tests. In particular,
Fig.~\ref{fig2} shows that if the equation of state were as soft as
predicted by the simple Pandharipande model, binary-pulsar tests
quantitatively supersede solar-system ones in all the region $\beta_0
\lesssim -1$ of parameter space. Even if we consider the less
constraining stiff equations of states, the present work confirms the
limit
\begin{equation}
\beta_0 > -4.5 \label{eq5.1}
\end{equation}
found (modulo $10\%$) in \cite{def96}. We recall that this limit can
be interpreted as a limit on the ratio of the two weak-field
post-Einstein parameters
\begin{equation}
\frac{\beta_{\rm Edd} - 1}{\gamma_{\rm Edd} - 1}
\approx - \frac{1}{4} \,
\beta_0 < 1.1 \, . \label{eq5.2}
\end{equation}

Finally, we pointed out that the discovery of a binary pulsar with a
black-hole companion has the potential of providing a superb new
probe of relativistic gravity. The discriminating power of this
probe might supersede all its present and foreseeable competitors in
measuring $\alpha_0$ down to the level $\alpha_0^2 < 10^{-6}$.

\acknowledgments
We wish to thank P.~Haensel for providing us with tables of data
describing the realistic equations of state that we used in our
numerical calculations. We also thank Z.~Arzoumanian, B.~Datta,
K.~Nordtvedt, J.H.~Taylor, S.E.~Thorsett, and C.M.~Will for
informative exchanges of ideas.

\appendix
\section{Finite-size effects in tensor-scalar gravity}

In the text we have assumed that the leading modification, in
tensor-scalar gravity, of the orbital motion of binary systems comes
{}from the change in radiation reaction forces. In this appendix, we
briefly discuss the modifications of the orbital motion caused by the
finite extension of the bodies. Contrary to the pure spin 2 theory
where such effects are very small (because they are suppressed in
spherical bodies), the presence of a scalar field opens the possibility
of couplings to the spherical inertia moment $\int d^3 {\bf x} \, \rho
({\bf x}) \, {\bf x}^2$. In this appendix we use the general
diagrammatic approach of Ref.~\cite{def2pn} to confirm (and hopefully
better understand) the results of Nordtvedt
\cite{nordt83,nordt91,nordtvedt} based on some explicit calculations
valid only for weakly-self-gravitating extended bodies. The final
outcome is that finite-size effects can be neglected in a
matched-filter analysis of inspiralling compact binaries.

Following the indications given in \cite{def2pn}, we can formally, but
very generally, take extension effects into account by considering that
the effective action for each compact body is a {\it functional\/} of
the fields $\varphi$ and $g^*_{\mu\nu}$ which can be expanded
in terms of the values along some central worldline of their
spacetime derivatives (derivative expansion).
Namely, we take the following action for the
extended body (labeled $E$)
\begin{equation}
S_E^{\rm extended} = -c \int m_E^{\rm eff} \,
[\varphi , g_{\mu \nu}^*] \left(-
g_{\mu \nu}^* (z_E^{\lambda}) \, dz_E^{\mu} \,
dz_E^{\nu}\right)^{1/2}\, ,
\label{eqA1}
\end{equation}
with
\begin{eqnarray}
&& m_E^{\rm eff} \, [\varphi , g_{\mu \nu}^*] = m_E (\varphi (z)) + I_E
(\varphi) \, R^* + J_E (\varphi) \, u_E^{*\mu} \, u_E^{*\nu}
\, R_{\mu \nu}^* + K_E (\varphi) \, \Box^* \, \varphi \nonumber \\
&& + L_E (\varphi) \, u_E^{*\mu} \, u_E^{*\nu} \, \nabla_{\mu}^* \,
\partial_{\nu} \, \varphi + M_E (\varphi) \,
u_E^{*\mu} \, u_E^{*\nu} \,
\partial_{\mu} \, \varphi \, \partial_{\nu} \,
\varphi + N_E (\varphi) \,
g^{*\mu \nu} \, \partial_{\mu} \, \varphi \,
\partial_{\nu} \, \varphi \, .
\label{eqA2}
\end{eqnarray}

This is the most general action, expanded up to two derivatives of the
fields $\varphi$ and/or $g_{\mu \nu}^*$, for a body which is
spherically symmetric and static when unperturbed. [Terms which are
first order in derivatives are excluded by spherical symmetry or, for
$H_E (\varphi) \, u_E^{*\mu} \, \partial_{\mu} \, \varphi$, by
time-reversal symmetry.] This form, being generic, is valid for
strongly self-gravitating bodies. In this case, the $\varphi$-dependent
quantities $I_E$, $J_E$, $K_E$, $L_E$, $M_E$, $N_E$, define some
``scalar form factors'' of body $E$ which go beyond the basic effective
coupling $\alpha_E (\varphi) = \partial \ln m_E (\varphi) /
\partial \, \varphi$ associated with the point-like effective action
$S_E^{\rm point} = -c \int m_E (\varphi) \, d \, s_E^*$. [For
non-spherical and/or non-static bodies many other new scalar form
factors could, in principle, appear.]

Most of the {\it a priori\/} independent-looking terms in
Eq.~(\ref{eqA2}) can be easily shown either not to contribute at the
(observationally most relevant) first post-Keplerian level [i.e.,
$O(v^2 / c^2)$ beyond the Keplerian orbital motion], or to be
equivalent to other terms, modulo some redefinition of the dynamical
variables. Let us first recall that any correction term, in a
Lagrangian, which is proportional to the zeroth-order equations of
motion can be redefined away by shifting some of the dynamical
variables. Technically, $S_0 [\psi] + \epsilon \, I (\psi) \, \delta
S_0 / \delta \psi = S_0 [\psi'] + O (\epsilon^2)$ with $\psi' = \psi +
\epsilon \, I (\psi)$. In our case, the dynamical variables can be
either $g_{\mu \nu}^*$, $\varphi$ or $z_E^{\mu}$. In all cases, the
local redefinitions $\psi \rightarrow \psi'$ have no effects on the
observables at infinity (such as the periastron advance or the
evolution of the orbital phase of an inspiralling binary). Using such
local redefinitions of the dynamical observables (and formally
neglecting singular self-action terms, i.e., $\delta$-function
contributions to $m_E^{\rm eff} [\varphi]$), one easily checks the
following simplifications: The $J_E (\varphi)$ contribution can be
transformed (using the $g_{\mu \nu}^*$ field equations $R_{\mu \nu}^* =
2 \partial_{\mu} \, \varphi \, \partial_{\nu} \, \varphi +
\delta$-functions and a local redefinition of $g_{\mu \nu}^*$ within
body $E$) into the $M_E (\varphi)$ contribution. The $M_E (\varphi)$
term contributes only at the second post-Keplerian level $O (v^4 /
c^4)$. The $K_E (\varphi)$ term can be eliminated by redefining
$\varphi$ within body $E$. The $L_E (\varphi)$ term can be reabsorbed
in the $N_E (\varphi)$ one by redefining the position $z_E^{\mu}$, and
by neglecting second post-Keplerian (2PK) effects. Indeed, integrating
by parts $- \int cds_E^* \, L_E (\varphi) \, u^{*\mu} \, u^{*\nu} \,
\nabla_{\mu}^* \, \partial_{\nu} \, \varphi$ one finds an $M_E
(\varphi)$-like term (contributing only at 2PK order) plus a term
proportional to $u^{*\mu} \, \nabla_{\mu}^* \, u^*_{\nu}$, which is
equal to $-\alpha_E (\delta_{\nu}^{\mu} + u^{*\mu} \, u_{\nu}^*) \,
\partial_{\mu} \, \varphi$ upon using the equations of motion. This
shows that after the redefinition
\begin{equation}
z_E^{\mu} \rightarrow z_E'^{\mu} = z_E^{\mu} +
\frac{L_E}{\widetilde{m}_E} \ \nabla^{*\mu} \varphi \, ,
\label{eqA3}
\end{equation}
the $L_E (\varphi)$ contribution is, modulo 2PK, equivalent to changing
$N_E (\varphi) \rightarrow N'_E (\varphi) = N_E + \alpha_E \, L_E$.
Similarly, the $I_E$ term can be absorbed in the $N_E$ one by locally
redefining $g_{\mu \nu}^*$. [In fact, if one fixes the harmonic gauge
before making this redefinition of $g_{\mu\nu}^*$, one also needs a
shift of the positions of the bodies $A\neq E$ to absorb $I_E$ into
$N_E$, namely, $z_A^i \rightarrow z_A'^i = z_A^i + 4 G_* I_E
(z_A^i-z_E^i) / (r_{EA}^3 c^2) + O(1/c^4)$.]

Finally, we end up (modulo 2PK) with a generic, simplified effective
action containing only a $N_E$-type two derivative term:
\begin{equation}
S_E^{\rm new} = -c \int ds_E^* \, [m_E (\varphi) +
N_E^{\rm new} (\varphi) \,
g^{*\mu \nu} \, \partial_{\mu} \, \varphi \,
\partial_{\nu} \, \varphi ] \, ,
\label{eqA4}
\end{equation}
where
\begin{equation}
N_E^{\rm new} = N_E + \alpha_E \, L_E + 2I_E \, . \label{eqA5}
\end{equation}
The conclusion is that, modulo $O (v^4 / c^4)$ terms and higher-order
terms in the radius of the extended body $E$ (associated to
higher-derivatives), there is only one relativistic form factor for,
possibly compact, extended bodies in tensor-scalar gravity: $N_E^{\rm
new}$. By dimensional analysis $[N_E^{\rm new}] = [{\rm mass}] \, [{\rm
length}]^2$ and we therefore expects $N_E^{\rm new}$ to be, roughly,
some spherical inertia moment. Using the diagrammatic method of
Ref.~\cite{def2pn}, it is easy to compute the extra contribution to the
$N$-body Lagrangian entailed by the presence of $N_E$:
\begin{equation}
\delta^{\rm extended} \, L_{N{\rm -body}} = - \sum_{A \ne E \ne B} \
\frac{G_*^2 \, m_A \, m_B \, (\alpha_A \, N_E^{\rm new} \,
\alpha_B) \, {\bf
n}_{EA} \cdot {\bf n}_{EB}}{r_{EA}^2 \, r_{EB}^2\, c^2}
\, , \label{eqA6}
\end{equation}
where ${\bf n}_{EA} \equiv ({\bf z}_E-{\bf z}_A)/r_{EA}$. Note that in
a two-body system the summation will have two terms: one where $E = 1$,
$A = B = 2$, and one where $E = 2$, $A = B = 1$. [We denote here for
clarity the two bodies by the labels 1 and 2.]

The above considerations have the advantage of being valid even when
discussing strongly self-gravitating extended bodies. In the case of a
{\it weakly\/} self-gravitating extended body, Nordtvedt
\cite{nordtvedt} has obtained an effective action, after some explicit
calculations of the equations of motion, of the form \begin{equation}
S_E = -c \int ds_E^* \, \left[ m_E (\varphi) + \frac{1}{6} \
\widetilde{I}_E \, A^2 (\varphi) \, \widetilde{u}^{\mu} \,
\widetilde{u}^{\nu} \, \widetilde{R}_{\mu \nu} \right] \, ,
\label{eqA7} \end{equation} where $\widetilde{I}_E = \int d^3
\widetilde x \, \widetilde \rho (\widetilde x ) \, \widetilde{\bf x}^2$
is the physical spherical inertia moment, and where we recall that the
tilde refers to physical, Jordan-frame quantities: $\widetilde{g}_{\mu
\nu} = A^2 \, g_{\mu \nu}^*$, $\widetilde{R}_{\mu \nu} = R_{\mu \nu}
[\widetilde g ]$, $\widetilde{g}_{\mu \nu} \, \widetilde{u}^{\mu} \,
\widetilde{u}^{\nu} = -1$. By expanding (\ref{eqA7}) in terms of our
generic expansion (\ref{eqA2}) we find
\begin{eqnarray}
I_E = 0 \ , \ J_E = \frac{1}{6} \, \widetilde{I}_E \ , \ K_E =
\frac{\alpha_0}{6} \, \widetilde{I}_E \ , \ L_E =
- \frac{\alpha_0}{3} \,
\widetilde{I}_E \nonumber \\
M_E = -\frac{1}{3} \, (\beta_0 - \alpha_0^2) \,
\widetilde{I}_E \ , \ N_E^{\rm
old} = \frac{1}{6} \, (\beta_0 + 2 \alpha_0^2) \, \widetilde{I}_E \, .
\label{eqA8}
\end{eqnarray}
We conclude from Eq.~(\ref{eqA5}) that the only observable form factor
{}from infinity is
\begin{equation}
N_E^{\rm new} = \frac{1}{6} \, \beta_0 \,
\widetilde{I}_E \, . \label{eqA9}
\end{equation}
In other words, after the shift $z_E'^{\mu} = z_E^{\mu} - \frac{1}{3}
\, \alpha_0 \, (\widetilde{I}_E / \widetilde{m}_E) \, \nabla^{*\mu} \,
\varphi$, the only new contribution to the $N$-body Lagrangian is
Eq.~(\ref{eqA6}) with $N_E^{\rm new}$ given by (\ref{eqA9}). When
comparing with Nordtvedt's results, note that he does not perform the
simplifying shift that we are advocating so that he ends up with a more
complicated-looking $N$-body Lagrangian containing a contribution like
(\ref{eqA6}) but with $N_E^{\rm old} = \frac{1}{6} \, (\beta_0 + 2
\alpha_0^2) \, \widetilde{I}_E$, plus two other sums which are simply
obtainable by applying the inverse shift ${\bf z}_E^{\rm Nordtvedt} =
{\bf z}_E^{\rm us} + \frac{1}{3} \, \alpha_0 \, (\widetilde{I}_E /
\widetilde{m}_E)\, \mbox{\boldmath$\nabla$} \varphi$, with
$\varphi = \varphi_0 - \sum_A G_* \, m_A \, \alpha_A / r_A \, c^2$,
to the unperturbed Lagrangian $L_0 ({\bf z}_A , {\bf v}_A) = \sum_A
\frac{1}{2} \, m_A \, {\bf v}_A^2 + \sum_{A \ne B} \frac{1}{2} \, G_*
\, m_A \, m_B \, (1+\alpha_A \, \alpha_B) / r_{AB}$. [Here, both
${\bf v}_A = \dot{\bf z}_A$ and $r_{AB} = \vert {\bf z}_A - {\bf z}_B
\vert$ get modified by the shift.] Finally, we find that in a binary
system the finite-size effects cause the appearance of the additional
interaction energy terms
\begin{equation}
\delta^{\rm extended} \, E_{\rm int} =
+ \frac{G_*^2 \, m_A \, m_B}{r_{AB}^4\, c^2}
\ (\alpha_A^2 \, N_B^{\rm new}
+ \alpha_B^2 \, N_A^{\rm new}) \, .
\label{eqA10}
\end{equation}
For compact bodies (radius $R_A$ comparable to $G_* \, m_A / c^2$) we
expect that $N_A^{\rm new} \sim \beta_A \, m_A (G_* \, m_A / c^2)^2$,
so that the new interaction energy (\ref{eqA10}) will be of the form
\begin{equation}
\delta^{\rm extended} \, E_{\rm int} \sim
(C_B \, \beta_B \, \alpha_A^2 + C_A \, \beta_A \,
\alpha_B^2) \, E_{\rm 3PK}^{\rm GR} \, , \label{eqA11}
\end{equation}
where $E_{\rm 3PK}^{\rm GR}$ denotes the general relativistic third
post-Keplerian $(O (v^6 / c^6))$ contribution to the interaction
energy, and where $C_A$ and $C_B$ are numerical coefficients which are
roughly of order unity. The 3PK energy $E_{\rm 3PK}^{\rm GR} \sim G_*
\, m_A \, m_B / r_{AB} \times (G_* \, m_A / c^2 \, r_{AB})^3$ depends
on the distance $r_{AB}$ in the same way as $\delta^{\rm extended} \,
E_{\rm int}$.

The conclusion is that scalar-mediated finite-size effects can be
neglected in a matched-filter analysis of the phase evolution of
inspiralling binaries, because they only modify by a factor $\sim 1 +
C\, \beta \, \alpha^2$ (which tends to 1 as $\alpha^2 \rightarrow 0$)
terms already present in the general relativistic phase evolution. As
discussed for similar fractional corrections in Section III, they can
be neglected compared to the non-general-relativistic scalar dipolar
contribution to the phase evolution.

\end{document}